\documentclass[review,sort&compress]{elsarticle}

\usepackage[pagewise]{lineno}
\modulolinenumbers[1]
\usepackage{hyperref}
\usepackage{svg}
\hypersetup{colorlinks=true,pdfborder=0 0 0}
\usepackage{subcaption}
\usepackage{amsmath,bm,siunitx,amssymb,caption,booktabs}

\usepackage{textcomp,gensymb}
\usepackage{cleveref}
\usepackage{ulem}
\usepackage{xcolor}
\journal{arXiv}

\begin{document}

\begin{frontmatter}

\title{AnisoGNN: graph neural networks generalizing to anisotropic properties of polycrystals}

\address[az-mse]{Department of Materials Science and Engineering, University of Arizona, Tucson, AZ 85721, USA}
\address[az-am]{Graduate Interdisciplinary Program in Applied Mathematics, University of Arizona, Tucson, AZ 85721, USA}

\author[az-mse]{Guangyu Hu}
\author[az-mse,az-am]{Marat I. Latypov\corref{cor1}}
\cortext[cor1]{corresponding author}
\ead{latmarat@arizona.edu}

\begin{abstract}

 We present AnisoGNNs -- graph neural networks (GNNs) that generalize predictions of anisotropic properties of polycrystals in arbitrary testing directions without the need in excessive training data. To this end, we develop GNNs with a physics-inspired combination of node attributes and aggregation function. We demonstrate the excellent generalization capabilities of AnisoGNNs in predicting anisotropic elastic and inelastic properties of two alloys.

\end{abstract}

\begin{keyword}
Graph neural networks \sep Polycrystals \sep Computational homogenization \sep Mechanical properties  
\end{keyword}

\end{frontmatter}


\section{Introduction} 
\label{sec:intro}

Structural materials (metals, alloys, ceramics) essential across multiple industries are predominantly used in their polycrystalline form constituted by numerous crystals known as {\it{grains}}. An interplay between the three-dimensional (3D) structure of grains and grain-level ({\it{single-crystal}}) properties dictate the macroscopic engineering performance of polycrystalline materials \cite{adams1998mesostructure}. 
Single-crystal properties of metals and alloys are often directionally dependent, i.e., {\it{anisotropic}}. This grain-level, microscale anisotropy together with non-random distribution of grain orientations ({\it{texture}}) result in a macroscopic anisotropy of overall (or {\it{effective}}) properties of polycrystals \cite{kocks2000texture}. For example, most metals and alloys have anisotropic single-crystal elastic properties described by a stiffness or compliance tensors \cite{nye1985physical}. A polycrystalline alloy with single-crystal anisotropy and non-random texture has therefore different values of the effective Young's modulus if tested in different {\it{sample directions}} (i.e., in the {\it{global frame}}). Similarly, most alloys, beyond the elastic limit, deform plastically by dislocation glide on discrete slip systems, which results in the macroscopic plastic anisotropy \cite{asaro1985overview}. Non-random textures are ubiquitous in alloys as they arise as a result of thermomechanical processing (e.g., casting, rolling, annealing) \cite{kocks2000texture}. Accordingly, the macroscopic elastic and plastic anisotropies are common in commercial alloys and have significant effects on their behavior during processing, manufacturing, and service. As a classical example, ``earing'' during cup drawing of aluminum alloys is a manifestation of mechanical anisotropy of practical significance: ears are processing defects that need trimming, leading to material losses \cite{engler2007polycrystal}. 

Models and simulations that predict effective anisotropic behavior of polycrystalline materials as a function of their microstructure are essential for optimization of industrial processing and in-service performance of alloys. Machine learning (ML) is emerging as a powerful computational framework for establishing quantitative microstructure--property relationships \cite{kalidindi2022digital}. ML models trained on data from numerical simulations offer an attractive combination of low computational cost and account for microstructure effects on  properties \cite{Latypov2017,Latypov2019,Zapiain2019,Pokharel2021,Paulson2017}. However, most ML models published to date do not systematically capture anisotropy of polycrystals. Capturing anisotropy in ML models is challenged by the need in training data for many sample directions. Generation of training data is time-consuming and expensive even with numerical simulations, not to mention experiments. This is why most ML models predict properties in only one (or three, at most) sample directions \cite{Paulson2017,hestroffer2023graph,Dai2021,dai2023graph}. One of the few published studies to date that capture anisotropic behavior of polycrystals relied on a large training dataset generated using a virtual laboratory framework \cite{nascimento2023machine}. Two other related studies \cite{parkDevelopmentDeepLearning2023,deocazapiainPredictingPlasticAnisotropy2022} developed ML models to predict anisotropic properties of polycrystals from crystallographic textures by generating large training datasets with simulations in many sample directions. In this contribution, we present a new ML strategy that generalizes property predictions to arbitrary sample directions {\it{without the need in excessive training data thereby reducing the computational cost}} of the ML model development. We build our new strategy on graph neural networks (GNNs) as the base ML framework. 

Graph neural networks (GNNs) are rapidly emerging as a powerful ML approach to modeling polycrystals. GNNs rely on the graph representation of polycrystals, in which graph nodes represent grains, while graph edges connect nodes corresponding to neighboring grains that share a boundary \cite{Lenthe2020,vlassis2020geometric}. Such graphs offer a reduced-order representation of polycrystals that is compact (as opposed to full-field 3D data), yet captures 3D grain connectivity. Capturing connectivity and thus local grain neighborhood is essential for modeling properties determined by local interactions of neighboring grains \cite{lebensohn1993self}. The graph representation further allows incorporating rich information about the microstructure beyond grain connectivity. Properties of grains, grain boundaries, and the whole polycrystal can be incorporated as node-, edge-, and graph-level attributes. For example, node attributes can include grain centroid coordinates \cite{dai2023graph}, crystallographic orientation \cite{dai2023graph,hestroffer2023graph,vlassis2020geometric}, metrics of the grain size and shape \cite{hestroffer2023graph,Dai2021,vlassis2020geometric}, as well as local properties of individual grains \cite{dai2023graph,pagan2022graph}. Graph-level attributes can in turn represent effective properties of polycrystals, which are often of interest for learning and inference. 

Prior related work includes the following studies. Hestroffer et al.\ \cite{hestroffer2023graph} trained a GNN on crystal plasticity finite element (CPFE) simulation data to predict the yield strength and Young's modulus of titanium in a single loading direction. Pagan et al.\ \cite{pagan2022graph} presented a GNN model of the elastic stress response of nickel and titanium alloys in one loading direction based on CPFE and 3D X-ray diffraction data. Dai et al.\ \cite{Dai2021} trained a GNN model to phase field data to predict effective magnetostriction in one sample direction under applied magnetic field. Dai et al.\ \cite{dai2023graph} modeled effective ion conductivities and elastic moduli in three principal sample directions by GNNs trained on numerical data from the Fourier spectral iterative perturbation method. Vlassis et al.\ \cite{vlassis2020geometric} developed a GNN that predicts hyperelastic energy functional from which mechanical responses of polycrystals can be derived. 

\section{AnisoGNN: new GNN framework for anisotropic properties}

Here, we present \texttt{AnisoGNN} -- a new GNN strategy to capture anisotropic properties in a wide range of directions {\it{without the need in training data in those directions}}. Our framework relies on fundamental symmetry properties of tensors and crystallographic orientations of individual grains and polycrystals. Specifically, we propose and implement two innovative features in GNN models for polycrystals: (i) microstructure rotation for training and inference and (ii) tensor properties as node attributes with physics-inspired aggregation.

\begin{figure}[!ht]
\centering
\includegraphics[width=\linewidth]{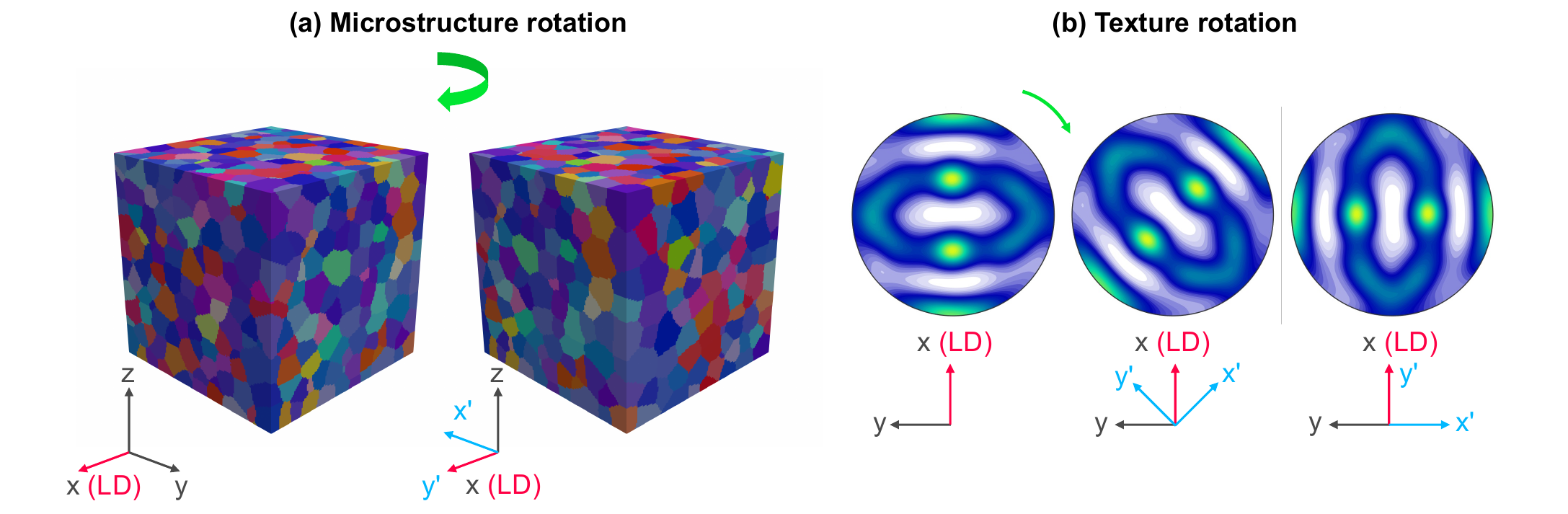}
\caption{Illustration of how effective properties in different directions can be probed along a fixed loading direction (LD) by rotation of the (a) microstructure, (b) texture.}
\label{fig:rot}
\end{figure}

With {\it{microstructure rotation}}, we emulate obtaining effective properties in different sample directions by rotating the microstructure in respect to the global frame. That is, if we are to obtain effective properties in $N$ sample directions, rather than simulate the material along these directions, we rotate the microstructure $N$ times and probe the property of interest along a fixed sample direction. For example, an effective property in the $x$ direction of a microstructure rotated \SI{90}{\degree} around the $z$ axis is equivalent to the effective property in the $y$ direction (\Cref{fig:rot}a). At the same time, to a GNN model, the rotated microstructure appears as yet another microstructure for inference. While the difference is subtle, modeling effective properties in a fixed sample direction 
is beneficial because it requires, as we shall show, less training data compared to ML of properties in multiple directions relying on data for each individual direction. For equiaxed polycrystals, it may suffice to rotate only texture, e.g., grain orientations in the microstructure (\Cref{fig:rot}b). 

{\it{Tensor node attributes with physics-inspired aggregation.}} We further implement most relevant tensorial properties of grains as node attributes in polycrystal graphs together with a physics-inspired aggregation function in GNNs. While prior studies used the crystallographic orientation as a key attribute for grain-nodes \cite{hestroffer2023graph,pagan2022graph,Dai2021,dai2023graph,vlassis2020geometric}, crystallographic orientation is not necessarily most compatible with aggregator functions commonly used in GNNs. Consider a cluster of neighboring grains represented by a set of graph nodes connected by graph edges. A mean or max aggregator function \cite{sanchez2021gentle} applied to Euler angles (or other orientation representation, e.g., quaternions) of these graph nodes would result in a new crystallographic orientation that is not representative of any of the grains in the cluster. In contrast, our AnisoGNN framework implements a grain-level tensor property most relevant to the effective property of interest. For example, for effective {\it{elastic}} properties, e.g., Young's modulus, we consider the local stiffness tensor of a grain as the most relevant set of attributes for the corresponding node. For modeling an effective {\it{inelastic}} property, we consider the local Schmid tensors of a grain as the most relevant set of attributes for the corresponding node. These property tensors are typically known at the grain level as single-crystal elastic constants in the case of the stiffness tensor and as a tensor product of the slip plane normal and slip direction vectors in the case of the Schmid tensor \cite{kalidindi1992crystallographic}. Since the Schmid tensor is defined for each slip system, we propose to concatenate elements of all tensors into a single vector, which will contain $9\cdot M$ elements for a metal or alloy with $M$ slip systems. Relevant single-crystal tensors can be identified for other (e.g., non-mechanical) effective properties as well. Upon selecting the relevant single-crystal tensor, we rotate it from the crystal frame to the global frame using the crystallographic orientation for each grain. These tensorial properties of grains rotated to the global frame collectively constitute the response of the polycrystal upon loading (or other stimulus) and thus serve as excellent candidates for attributes of grain-nodes. With these attributes, a simple mean aggregator function often used in GNNs \cite{sanchez2021gentle} gains a physical meaning -- the mean of the tensor properties of the neighboring grains in a cluster represents their average tensor property in the macroscopic frame, which is consistent with simple micromechanics models of polycrystals \cite{diz1992practical}. 

With tensor properties as graph-node attributes, we design a simple GNN architecture with a graph convolutional layer implementing mean aggregation function followed by global mean pooling and a final fully-connected layer for output -- polycrystal's effective property of interest.

\section{Case studies}
\label{sec:case}

We demonstrate and critically evaluate our new AnisoGNN framework in two case studies: (i) effective anisotropic elastic properties of a polycrystalline Ni-base superalloy; (ii) effective anisotropic elastic and plastic properties of polycrystalline aluminum. For both case studies, we used a set of 300 digitally generated 3D microstructure volume elements (MVEs, as described in \Cref{sec:micro}). The MVEs represented 3D equiaxed microstructures with 12 distinct {\it{initial}} textures. Four initial textures were deformation textures after uniaxial tension, uniaxial compression, plain strain compression, and simple shear loading. To generalize the AnisoGNN models to different loading directions, the microstructure dataset further included MVEs with rigidly rotated instances of these four textures. Four textures were obtained by rotation to \SI{45}{\degree} and four additional textures were obtained by rotation to \SI{90}{\degree} -- both about the $z$ axis. We used orientation distribution functions for these 12 textures to sample crystallographic orientations and randomly assign them to grains of 25 MVEs generated for each texture. We used the graphs representing the MVEs and MVEs' effective properties from simulations to train AnisoGNN models. We tested the ability of AnisoGNN models to predict anisotropic properties under four distinct scenarios: (i) random split of the 300 dataset with \SI{70}{\percent} used for training and \SI{30}{\percent} used for testing, (ii) 100 MVEs with \SI{45}{\degree} rotated textures used for testing, (iii) 100 MVEs with \SI{90}{\degree} rotated textures used for testing, (iv) 200 MVEs with both \SI{45}{\degree} and \SI{90}{\degree} rotated textures used for testing. For all four scenarios, we trained two GNN models: (a) a benchmark GNN with previously published architecture \cite{hestroffer2023graph} utilizing SAGE convolution \cite{Hamilton2017} and crystallographic orientations as graph node attributes, as well as (b) new AnisoGNN with simple architecture that uses mean aggregation function and elements of a relevant property tensor as graph node attributes. \Cref{sec:gnn} provides further details of the GNNs. 

\subsection{Elastic properties of Ni-base superalloy}
\label{sec:ni}

Our first case study focused on modeling the effective elastic modulus of a Ni-base superalloy Ren{\'e} 88DT. Ren{\'e} 88DT has a relatively high single-crystal elastic anisotropy (among cubic metals) with the elastic constants $C_{11}=267.1$ \SI{}{\giga\pascal}, $C_{12}=170.5$ \SI{}{\giga\pascal}, and $C_{44}=107.6$ \SI{}{\giga\pascal} \cite{du2018first} and the corresponding anisotropy ratio of $A_c=2.2$ \cite{latypov2021insight}. To generate data for training the GNN models, we obtained effective elastic modulus values for 300 MVEs with crystal elasticity numerical simulations (see details in \Cref{sec:fft}). As noted above, we trained two models: (i) ``SAGE-O'' model trained on graphs with crystallographic orientations as grain-node attributes, and (ii) ``AnisoGNN-C'' model trained on graphs with local stiffness tensor as grain-node attributes. For the SAGE-O model's attributes of graph nodes, we used quaternions as the orientation description of choice. For the AnisoGNN-C model's graphs, each grain-node was assigned an attribute vector of the full stiffness tensor (defined in respect to the global frame), which describes the local elastic properties of the grain. To obtain the stiffness tensor for each grain in the global frame, we rotated the single crystal stiffness tensor (with $C_{11}$, $C_{12}$, and $C_{44}$ as the only non-zero components) from the crystal frame according to the grain's crystallographic orientation. Both the SAGE-O model and the AnisoGNN-C model had the grain volume (in voxels) as an additional attribute of all grain-nodes. We evaluated the performance of these two GNN models and their predictive abilities in the four training/testing scenarios listed above. 

\begin{figure}[t]
\centering
\includegraphics[width=\linewidth]{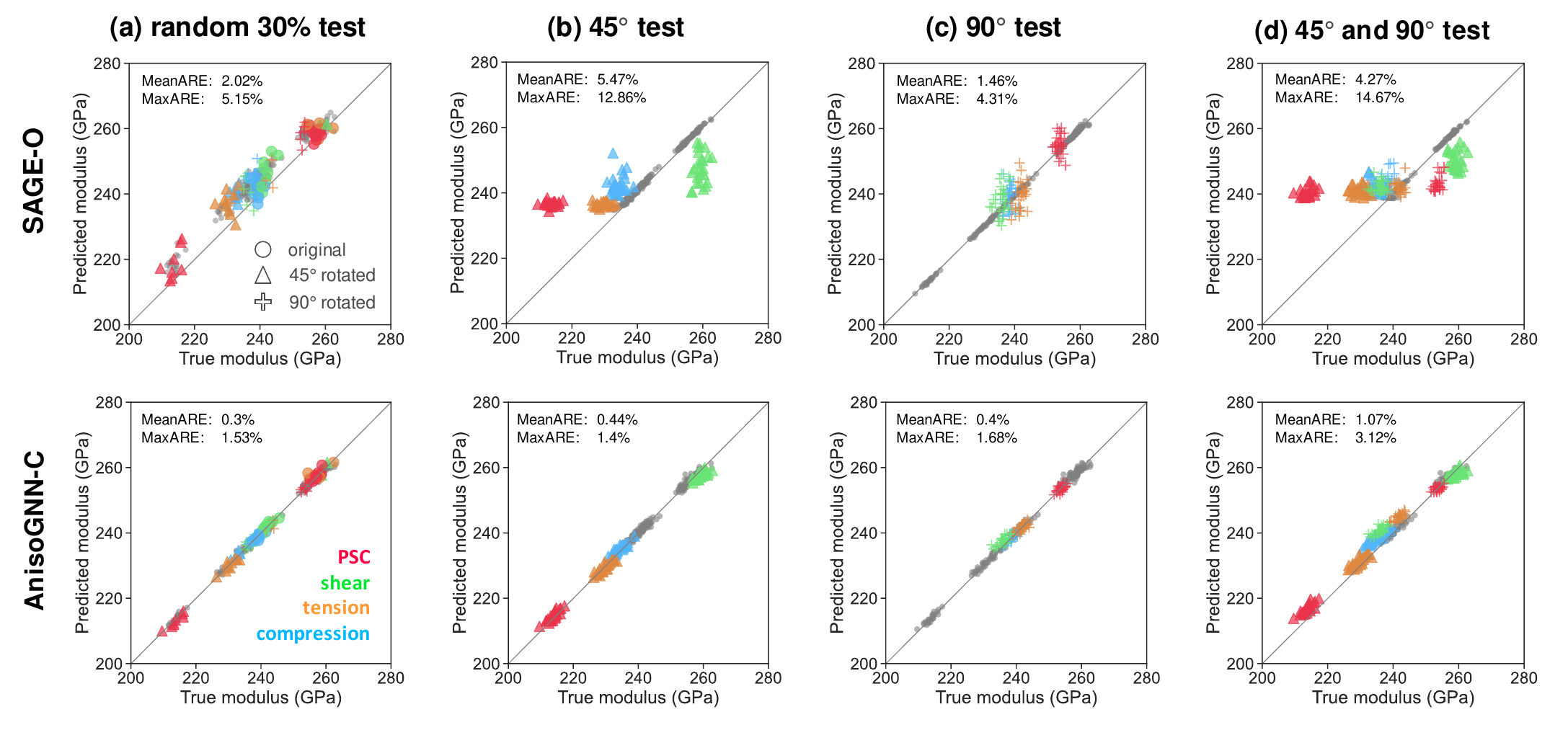}
\caption{GNN predictions of the effective elastic modulus of Ren{\'e} 88DT in the $x$ direction for different test sets of ``unseen'' MVEs: (a) 90 randomly picked MVEs; (b) MVEs with textures rotated \SI{45}{\degree} around the $z$ axis; (c) MVEs with textures rotated \SI{90}{\degree} around the $z$ axis. The predictions are depicted in parity plots against ground truth results obtained by the numerical simulations. The color of the symbols in the plots represents the texture type, while the shape represents the rotation of the textures.}
\label{fig:ni}
\end{figure}

\Cref{fig:ni} shows the performance of the two GNN models in predicting effective elastic modulus of the Ren{\'e} 88DT alloy for the test sets of MVEs unseen by the models during training. The predictions are visualized in parity plots against numerical simulation results considered as ground truth. The four parity plots for each GNN model correspond to the four training/testing scenarios described above. 

First of all, the AnisoGNN-C model outperforms the benchmark SAGE-O model in all four scenarios. Both GNN models demonstrate the highest accuracy when predicting the elastic modulus for the randomly selected subset of MVEs (\Cref{fig:ni}a). This outcome is expected because the GNN models are exposed to the MVEs representing all textures during training in the case of the random split of the MVE dataset. However, it is the predictive power of the models on rotated textures that is of most interest in the present study. Indeed, predicting the elastic modulus for a \SI{90}{\degree} rotated texture emulates predicting the modulus along the $y$ axis, while modulus for a \SI{45}{\degree} rotated texture corresponds to the modulus along the direction between the $x$ and $y$ axes. In this context, the AnisoGNN-C model exhibits excellent agreement with the ground truth values of the elastic modulus for MVEs of rotated textures. Remarkably, the AnisoGNN-C model accurately predicts the elastic modulus for MVEs of both rotated textures unseen during training: that is, even when the model is trained only on the unrotated textures (\Cref{fig:ni}d: 100 MVEs for training, 200 MVEs for testing). 

The superior generalization ability of our new AnisoGNN model to rotated textures is apparent when compared to the benchmark SAGE-O model. The SAGE-O model exhibits significantly lower accuracy when predicting the elastic modulus of MVEs with the rotated textures. Specifically, for the \SI{90}{\degree} rotated texture, the SAGE-O model demonstrates a wider variance around the ground truth values compared to the AnisoGNN-C model (\Cref{fig:ni}c). Further, the SAGE-O model significantly overestimates the modulus values for MVEs with a \SI{45}{\degree} rotated plane-strain compression texture (\Cref{fig:ni}b). The SAGE-O model also overestimates the modulus for the MVEs with the rotated uniaxial compression and tension textures, however to a lesser extent than plain-strain compression texture, and underestimates the modulus for MVEs with the shear texture. We observe similar trends of overestimation and underestimation by the SAGE-O model for the same textures in the training scenario where the MVEs of all rotated textures are excluded from training (\Cref{fig:ni}d).

The superior performance of the AnisoGNN-C model is corroborated by lower mean absolute relative errors (MeanARE, also displayed in \Cref{fig:ni}). The AnisoGNN-C model demonstrates a slightly better performance for the case of the \SI{90}{\degree} rotated texture as the test set than the SAGE-O model (\Cref{fig:ni}c) and three to ten-fold lower MeanARE values for other testing scenarios.

\subsection{Elastic and inelastic properties of aluminum}
\label{sec:al}

Our second case study focused on modeling the effective elastic modulus as well as the effective yield strength of aluminum. In contrast to nickel, aluminum and its alloys are characterized by low elastic anisotropy ($A_c=1.2$) with the following typical elastic constants: $C_{11}=106.75$ \SI{}{\giga\pascal}, $C_{12}=60.41$ \SI{}{\giga\pascal}, $C_{44}=28.34$ \SI{}{\giga\pascal} \cite{eisenlohr2013spectral}. The development and evaluation of the GNN models for the elastic modulus of aluminum were identical to those described for Ren{\'e} 88DT. That is, we trained similar SAGE-O and AnisoGNN-C models on data from numerical elasto-plasticity simulations on polycrystalline MVEs and evaluated them in the same four training/testing scenarios. 

We used the results from the same simulations for training GNN models targeting the effective yield strength. 
As noted above, in our AnisoGNN approach, we propose to use the Schmid tensors as node attributes in polycrystal graphs for modeling inelastic effective properties. We choose the Schmid tensors as the most relevant grain-level tensor properties because they capture the geometry of slip and critical resolved shear stresses of slip systems in individual grains \cite{asaro1985overview} that collectively define the overall yield strength of a polycrystal. As in the case of the stiffness tensor, we rotate the Schmid tensors of each grain to the global frame using the grain orientation information. For each grain, we rotate 12 Schmid tensors corresponding to the 12 slip systems in f.c.c.\ metals \cite{kalidindi1992crystallographic} and concatenate the elements of the rotated Schmid tensors (described by $3\times 3$ matrices) into a vector of 108 attributes.  We denote GNNs utilizing the Schmid tensor for the grain-node attributes as the AnisoGNN-S model, which we train and test in the four scenarios and compare their performance against the benchmark SAGE-O models that use orientation quaternions as grain-node attributes. 

\begin{figure}[t]
\centering
\includegraphics[width=\linewidth]{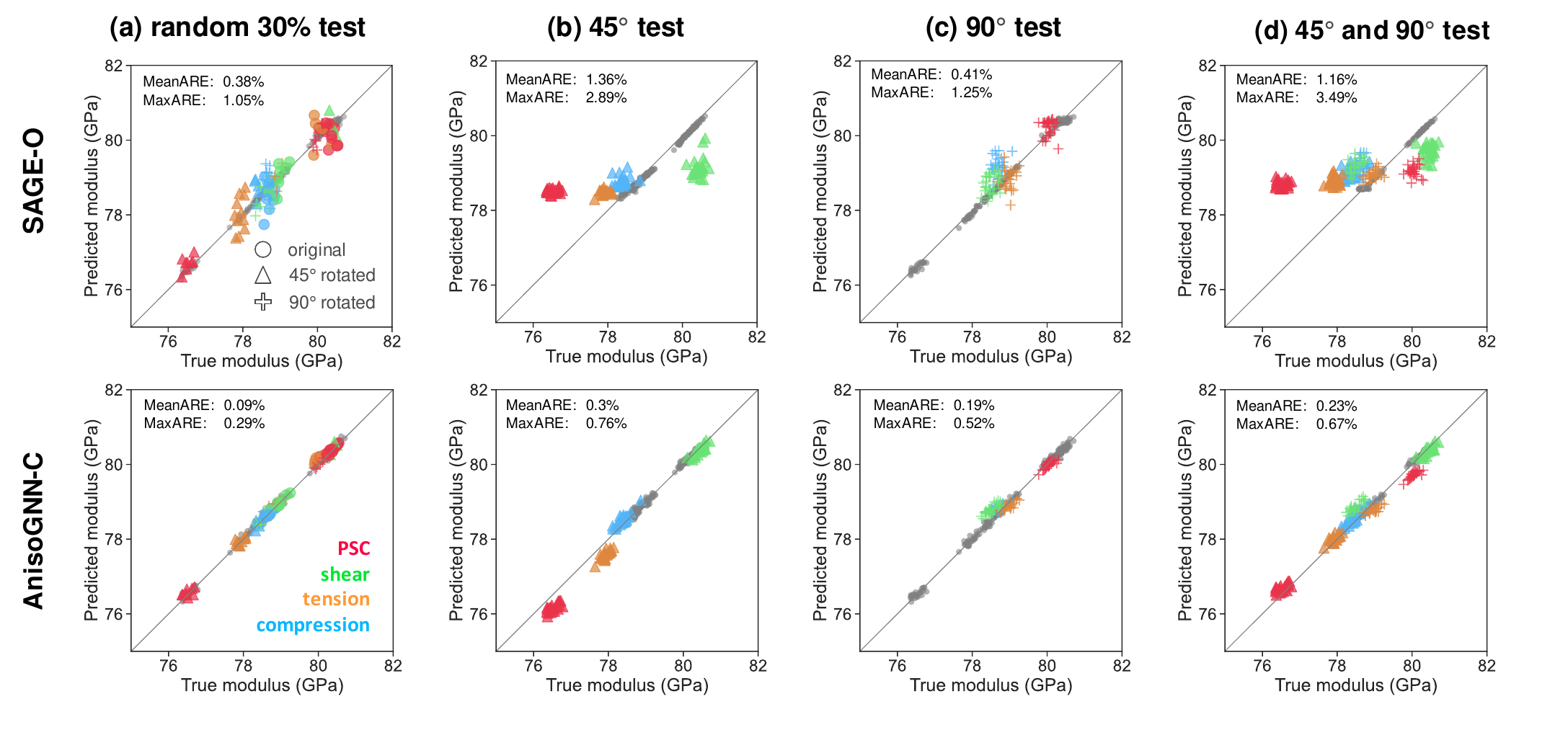}
\caption{GNN predictions of the effective elastic modulus of aluminum in the $x$ direction for different test sets of ``unseen'' MVEs: (a) 90 randomly picked MVEs; (b) MVEs with textures rotated \SI{45}{\degree} around the $z$ axis; (c) MVEs with textures rotated \SI{90}{\degree} around the $z$ axis. The predictions are depicted in parity plots against ground truth results obtained by the numerical simulations. The color of the symbols in the plots represents the texture type, while the shape represents the rotation of the textures.}
\label{fig:al-elas}
\end{figure}

The parity plots for the GNN models predicting the effective elastic modulus of aluminum polycrystals (\Cref{fig:al-elas}) show the same trends as those for the GNN models of Ren{\'e} 88DT. The AnisoGNN-C model exhibits better accuracy than the SAGE-O model in all four training/testing scenarios. The SAGE-O model shows good accuracy (MeanARE below \SI{0.4}{\percent}) for the randomly picked subset of MVEs (\Cref{fig:al-elas}a) and for the MVEs with \SI{90}{\degree} rotated texture (\Cref{fig:al-elas}c) but generalizes poorly for the other two test cases. At the same time, the AnisoGNN-C model maintains accuracy (MeanARE below \SI{0.3}{\percent}) across all testing cases, thereby demonstrating its strong generalization capabilities. 

\begin{figure}[t]
\centering
\includegraphics[width=\linewidth]{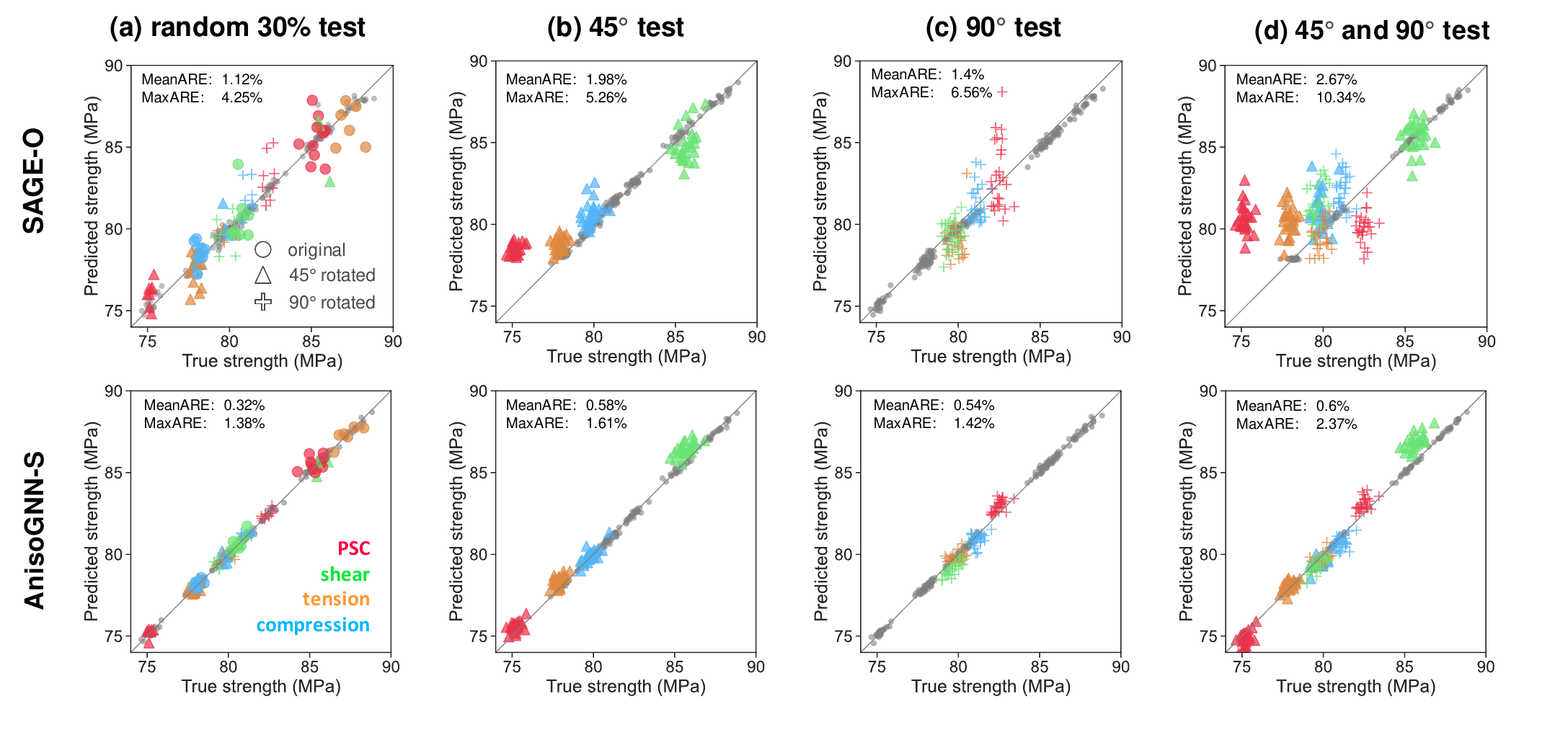}
\caption{GNN predictions of effective yield strength of aluminum in the $x$ direction for different test sets of ``unseen'' MVEs: (a) 90 randomly picked MVEs; (b) MVEs with textures rotated \SI{45}{\degree} around the $z$ axis; (c) MVEs with textures rotated \SI{90}{\degree} around the $z$ axis. The predictions are depicted in parity plots against ground truth results obtained by the numerical simulations. The color of the symbols in the plots represents the texture type, while the shape represents the rotation of the textures.}
\label{fig:al-plas}
\end{figure}

The results for the yield strength of aluminum (\Cref{fig:al-plas}) show that the generalizable predictive power of the AnisoGNN approach is not limited to elastic properties of polycrystals. The AnisoGNN-S model makes accurate predictions of yield strength of aluminum for all test sets of MVEs with the MeanARE values below \SI{0.6}{\percent}. The accuracy of the SAGE-O model, on the other hand, deteriorates when predicting yield strength for rotated textures (or similarly, predicting yield strength in new loading directions) not included in the training set. The agreement with the ground truth is worst when the SAGE-O model is trained only on MVEs of four unrotated textures (\Cref{fig:al-plas}d). Similar to modeling elastic modulus, the SAGE-O model tends to overstimate the yield strength for MVEs with plain-strain compression textures rotated \SI{45}{\degree} about the $z$ axis (see \Cref{fig:al-plas}b,d). 

\subsection{Effective properties in a wide range of directions}

\begin{figure}[!ht]
\centering
\includegraphics[width=0.7\textwidth]{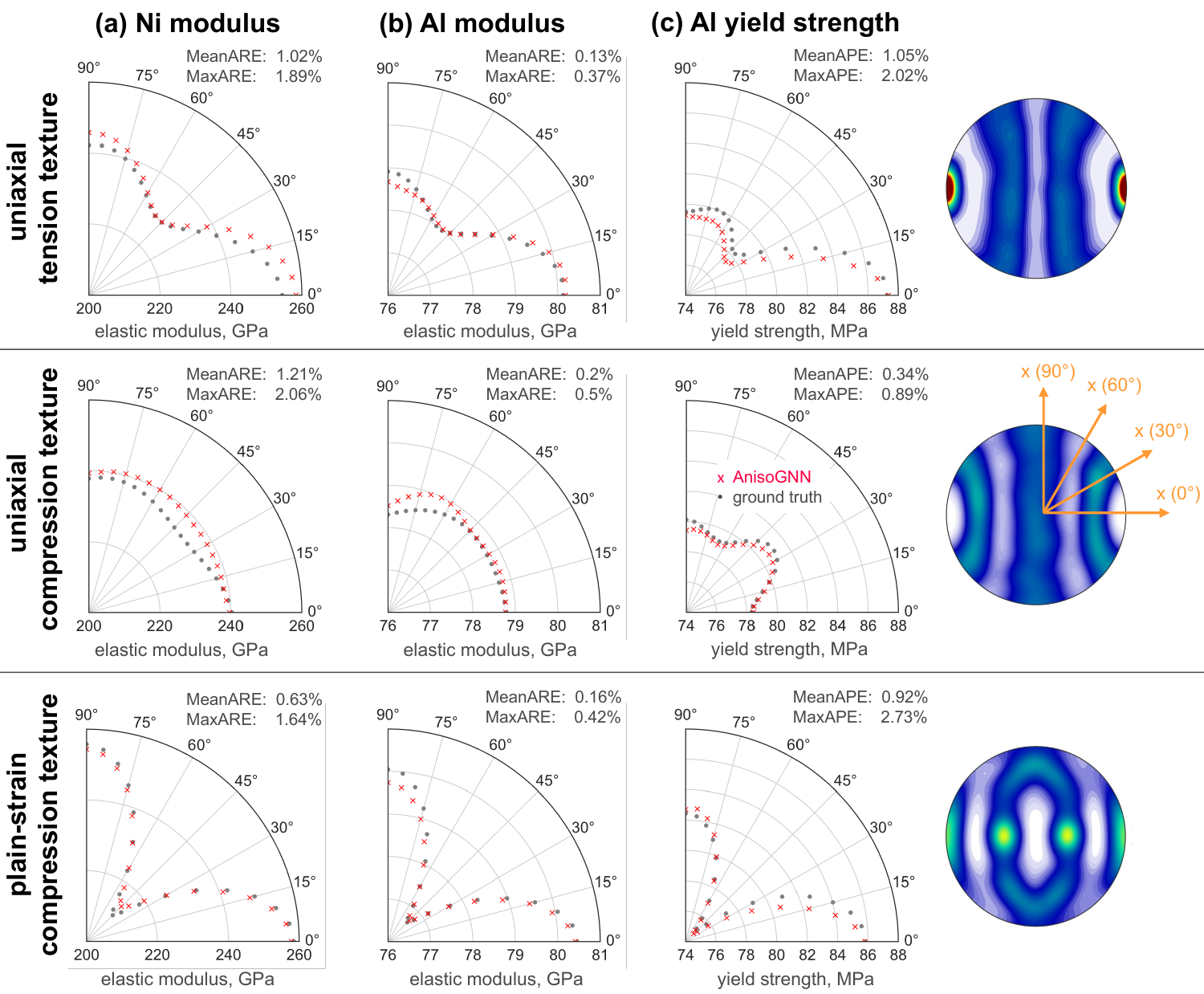}
\caption{GNN predictions of effective properties of Ren{\'e} 88DT and aluminum for MVEs of textures rotated to a range of angles.}
\label{fig:aniso}
\end{figure}

In the two case studies described above, we evaluated the predictive power of the AnisoGNN framework on MVEs with two sets of rotated textures, which emulates the prediction of effective properties in two new sample directions (along $y$ and along the direction between $x$ and $y$ axes). We finally show that the generalization of the AnisoGNN approach is not limited to only those two directions. \Cref{fig:aniso} compares predictions of the AnisoGNN models for three MVEs representative of three initial textures for a range of rotations around the $z$ axis with a \SI{5}{\degree} increment. The AnsisoGNN models generalize well for the range of rotations across all three initial textures and three alloy/property cases (elastic modulus of Ren{\'e} 88DT, elastic modulus and yield strength of aluminum). These models trained only on 100 MVEs with the unrotated initial textures predict these properties with meanARE not exceeding \SI{1.2}{\percent}.

\section{Discussion}

{\it{Grain tensor properties as node attributes}}. The flexibility of graphs to incorporate rich information about grains (geometric, crystallographic, and physical) is one of the powerful features of the GNN approach to modeling polycrystals. The use of crystallographic orientations as the key grain-node attributes is a natural choice \cite{hestroffer2023graph,pagan2022graph,Dai2021,dai2023graph,vlassis2020geometric}. Our results suggest, however, that the use of tensorial properties relevant to the target effective property significantly improves the accuracy and generalizing capabilities of the GNN models. Indeed, the GNN models predicted the elastic modulus with better accuracy when relied on grains' stiffness tensor rather than orientation quaternions for both Ren{\'e} 88DT with high elastic anisotropy and nearly isotropic aluminum (compare AnisoGNN-C and SAGE-O models in \Cref{fig:ni,fig:al-elas}). We observe the same for yield strength: the use of the Schmid tensor as grain-node attributes improves the accuracy and generalization of the GNN models as seen in our case study on aluminum (compare AnisoGNN-S and SAGE-O models in \Cref{fig:al-plas}). We attribute the superior performance of AnisoGNNs to the physical basis of the mean aggregation (common in GNNs \cite{sanchez2021gentle}) when used in combination with tensorial properties of the grains. That is, averaging elements of property tensors of individual grains in each grain cluster gives an estimate of the overall property of the cluster. Our AnisoGNN approach still accounts for crystallographic orientations because we use the orientation information to rotate tensor properties of each grain into the global frame.   

{\it{Crystallographic vs.\ morphological texture}}. We enabled modeling anisotropic properties of polycrystals by targeting effective properties in respect to a fixed direction in the global frame while rotating crystallographic orientations of the grains in MVEs. Strictly speaking, one needs to rotate the whole microstructure, not just the grain orientations for complete equivalency (\Cref{fig:rot}a). In this study, we focused on {\it{equiaxed}} polycrystalline microstructures, for which mechanical anisotropy is dictated mostly by crystallographic texture. Therefore, it suffices to rotate only textures while keeping the 3D spatial arrangement of grains stationary. For a quantitative confirmation, we ran a series of additional simulations and compared the elastic modulus of (the more anisotropic) Ren{\'e} 88DT in the $y$ axis of MVEs without texture rotation and the elastic modulus in the $x$ axis on the same MVEs with grain orientations
rotated \SI{90}{\degree} about the $z$ axis. The elastic modulus in the $y$ direction emulated by texture rotation agreed with the directly calculated values within \SI{0.6}{\percent} meanARE. The AnisoGNN-C model predictions capture the elastic modulus calculated explicitly along the $y$ direction within \SI{1.2}{\percent} meanARE. More generally, polycrystals may have non-equiaxed grains and morphological texture, e.g., preferred orientations of major axes of elongated grains. Anisotropic properties of non-equiaxed microstructures depend on both crystallographic and morphological textures. Our approach can be extended to microstructures with morphological texture by generating MVEs with not only rotated grain orientations but also rotated morphological textures (e.g., axes of non-equaxed grains) with tools like Dream.3D \cite{groeber2008framework}. 

{\it{Interpolation vs.\ extrapolation}}. It is well acknowledged that most ML methods work best in interpolation and may struggle to extrapolate predictions to ranges away from training data \cite{marcus2018deep}. When modeling anisotropic properties, one may expect that a GNN model can predict properties in directions between those included in the training dataset. Our results show that it is not necessarily the case. The SAGE-O models struggled to accurately predict effective properties for MVEs with textures rotated \SI{45}{\degree} even when the training dataset included textures rotated \SI{90}{\degree} and \SI{0}{\degree}  (i.e., unrotated) around the $z$ axis -- see case (b) in \Cref{fig:ni,fig:al-elas,fig:al-plas}. At the same time, the SAGE-O models captured well the effective properties for MVEs with texture rotated \SI{90}{\degree} after learning from MVEs with  \SI{0}{\degree} and \SI{45}{\degree} textures (see case (c) in \Cref{fig:ni,fig:al-elas,fig:al-plas}). Inspection of these two test cases suggests that models with limited generalization need training data covering a wide range of the {\it{target property}} rather than the rotations/directions. Indeed, the \SI{45}{\degree} test set requires GNN models to extrapolate to the range of low values of the effective modulus (210 to \SI{230}{\giga\pascal} in \Cref{fig:ni}b) not present in the training set (covering 235 to \SI{265}{\giga\pascal} in \Cref{fig:ni}b). In other words, the SAGE-O models can interpolate in respect to the effective property rather than texture rotations or sample directions. These limitations are lifted with the AnisoGNN models, which show excellent extrapolation capabilities for effective property values outside of the training sets -- see AnisoGNN results in cases (b) and (d) in \Cref{fig:ni,fig:al-elas,fig:al-plas}. The predictive power and generalization abilities of simple AnisoGNN-type models presented here can be potentially further improved by adopting neural network architectures implementing rotational equivarance \cite{thomas2018tensor,brandstetter2021geometric,geiger2022e3nn,deng2021vector}.

{\it{Computational cost}}. The simple architecture of the AnisoGNN models requires modest computational resources for training and inference. Training AnisoGNN-C models to 800 epochs on 200 MVEs (with approximately 1000 grains each) takes about \SI{160}{\second} on a computer with a consumer-grade GPU (AMD Ryzen 7 1700X 8-core CPU \SI{3.40}{\giga\hertz} with an NVIDIA GeForce GTX 1660 Ti). The AnisoGNN-S model takes slightly more time (\SI{200}{\second}) on the same computer and under the same training conditions (200 MVEs, 800 epochs) because of a larger attribute vector on each node: 108 elements of the Schmid tensor vs.\ 21 elements of the elastic tensor. Calculating properties with the trained AnisoGNN models (inference) is nearly instantaneous for hundreds of MVEs.  

\section{Summary}

In summary, we presented AnisoGNN -- a new GNN framework for modeling anisotropic elastic and inelastic properties of polycrystalline materials. Our GNN models generalize to properties in new sample directions by accurately predicting properties of rigidly rotated microstructures (textures in case of equiaxed grains). To ensure generalization abilities, we trained GNNs on graphs with physics-informed node attributes: tensorial properties of grains relevant to the effective property of interest. We obtain these attributes by rotating the single-crystal property tensors to the sample frame using the crystallographic orientation of each grain. We demonstrated the predictive power and generalization of our GNN models in two case studies: the elastic moduli of a Ni-base superalloy and both elastic moduli and directional-dependent yield strength of aluminum.

\section{Data availability}

The codes and data generated during the current study are available on GitHub at \href{https://github.com/materials-informatics-az/AnisoGNN}{https://github.com/materials-informatics-az/AnisoGNN}.


\section*{Acknowledgements}

The authors gratefully acknowledge the High Performance Computing resources supported by the University of Arizona TRIF, UITS, and the Office of Research, Innovation, and Impact, and maintained by the UArizona Research Technologies Department. The authors further thank Dr.\ Jonathan Hestroffer (University of California, Santa Barbara) for fruitful discussions on training graph neural networks.



\setcounter{section}{0}
\renewcommand{\thesection}{\Alph{section}}
\renewcommand{\theHsection}{appendixsection.\Alph{section}}

\section{Appendix: methods}
\label{sec:appendix-a}

\subsection{Microstructure generation with Dream.3D}
\label{sec:micro}

To build a training dataset for the new GNN model, we first generated digital 3D MVEs. Our procedure of the MVE generation included two separate steps of creating (i) 3D grains (grain structure) and (ii) their crystallographic orientations (texture). 

{\it{Grain structure}}. We used open-source software Dream.3D \cite{Groeber2014} to generate single-phase polycrystalline MVEs with $128\times128\times128$ voxels. The Dream.3D algorithm populates grains as ellipsoids and then grows them to fill the 3D space allocated for each MVE \cite{groeber2008framework}. When populating grains, Dream.3D matches user-defined statistics of the grain size and morphology. The grain size is described by a log-normal distribution of the equivalent sphere diameter (ESD). We adopted the values of $\mu=2.7$ for the mean and $\sigma=0.2$ for the standard deviation of ESD (in arbitrary units). 

The grain morphology is described by aspect ratios of the ellipsoid precursors of the grains. In this study, we targeted equiaxed microstructures and thus set the mean values of the aspect ratio distributions to 1. Using these settings for the grain size and morphology, we generated 300 MVEs. Each $128\times128\times128$ MVE (with each voxel \SI{1}{\micro \cubic m}) contained more than 1000 grains. For each grain in an MVE, we assigned crystallographic orientations sampled from separately generated orientation distribution functions (ODFs). 

\begin{figure}[!ht]
\centering
\includegraphics[width=0.9\linewidth]{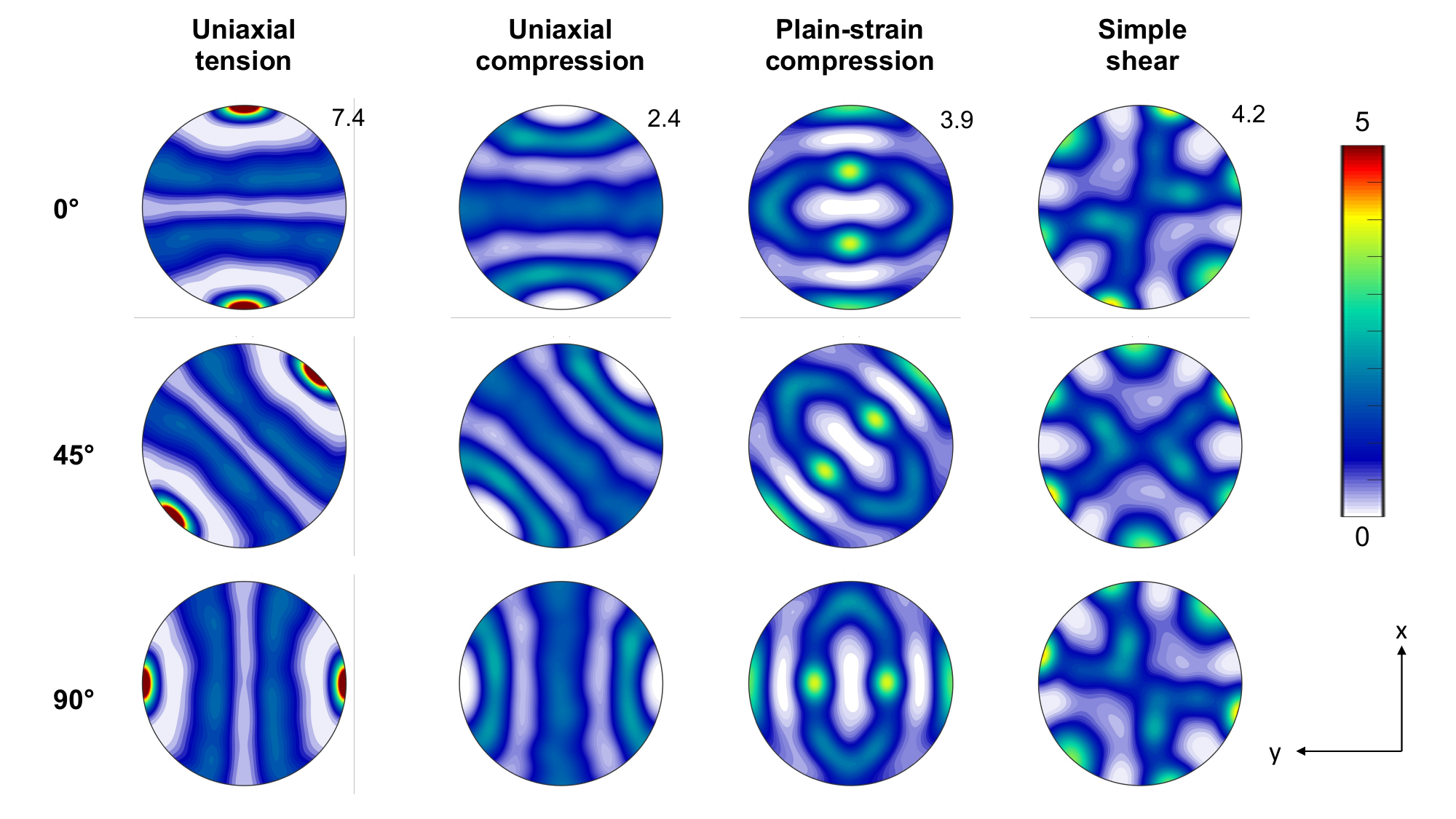}
\caption{Pole figures (for (111) plane normals) showing 12 textures used for microstructure generation in the present study. The angle in degrees indicates the rigid rotation around the $z$ axis (out of the figure plane). The number next to pole figures indicates the maximum intensity.}
\label{fig:pf}
\end{figure}

{\it{Crystallographic texture}}. We created multiple initial textures for MVEs to capture a diverse set of anisotropic mechanical properties. Textures were obtained by polycrystal plasticity simulations under different boundary conditions. We obtained four textures corresponding to uniaxial tension, uniaxial compression, plain strain compression, and simple shear deformation described by the following strain tensors: 



\begin{subequations}\label{eqn:comp}
\begin{align}
\mathbf{E}_\text{uniaxial} = \begin{pmatrix}
-\epsilon_\text{uni} &  0&  0\\
0 & \epsilon_\text{uni}/2 & 0 \\
0 & 0 & \epsilon_\text{uni}/2 \\ 
\end{pmatrix},
\end{align}

\begin{align}
\textbf{E}_\text{psc} = \begin{pmatrix}
\epsilon_\text{psc} &  0&  0\\
0 & 0 & 0 \\
 0& 0 & -\epsilon_\text{psc} \\
\end{pmatrix},
\end{align}

\begin{align}
\textbf{E}_\text{shear} = \begin{pmatrix}
0 &  \gamma &  0\\
0 & 0 & 0 \\
0 & 0 & 0 \\
\end{pmatrix},
\end{align}

\end{subequations}

\noindent where $\epsilon_\text{uni} = 1.1547$  for uniaxial tension,  $\epsilon_\text{uni} = -1.1547$ for uniaxial compression, $\epsilon_\text{psc} = 1$, and $\gamma=2$. These values were selected to result in the same von Mises value of 1.1547 for all four deformation tensors. We ran polycrystal plasticity simulations with these boundary conditions on a polycrystal consisting of 500 initially random crystal orientations. Our polycrystal plasticity simulations used Taylor homogenization scheme implemented in the open-source MTEX code \cite{MTEX}.

\begin{figure}[!ht]
\centering
\includegraphics[width=\linewidth]{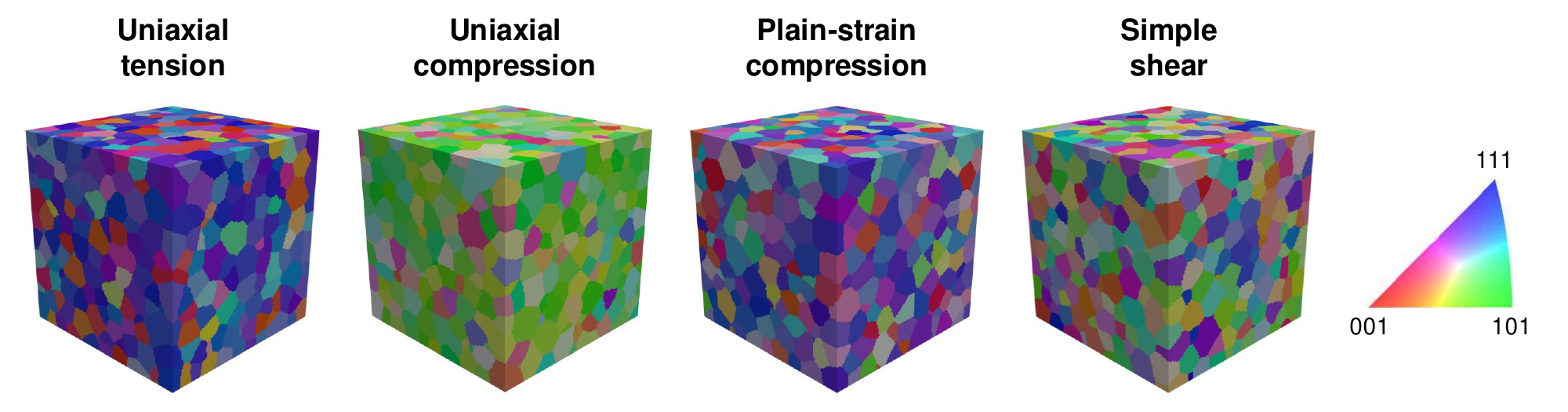}
\caption{Typical MVEs for each of the four unrotated textures generated in this study. Grains in the MVEs are color-coded according to their crystallographic orientations and using the inverse pole figure colors.}
\label{fig:mve}
\end{figure}

Following polycrystal plasticity simulations, we calculated ODFs from 500 discrete orientations of the deformed polycrystals. To further enrich the dataset, we obtained additional textures by rigid rotation of the four ODFs to \SI{45}{\degree} and \SI{90}{\degree} about the $z$ axis. We therefore obtained 12 distinct ODFs, including the textures obtained by polycrystal plasticity simulations and their two rotated versions (\Cref{fig:pf}). These 12 ODFs were used to sample discrete orientations for grains in the MVEs generated in Dream.3D as described above. For each of 12 textures, we allocated 25 MVEs with individual sets of discrete orientations sampled from the ODFs. The final microstructrue dataset contained 300 unique MVEs representing 12 initial textures (25 MVEs per texture) with more than 1000 grains in every MVE. \Cref{fig:mve} shows a few typical MVEs generated in this study. This microstructure dataset was used for micromechanical simulations of effective properties.


\subsection{Property simulations} 
\label{sec:fft}

To obtain effective properties of MVEs, we used fast Fourier transform (FFT) solver implemented within the Düsseldorf Advanced Material Simulation Kit (DAMASK) \cite{roters2019damask}. We performed FFT simulations of uniaxial tension by applying mixed boundary conditions described by the rate of the deformation gradient and the first Piola--Kichhoff stress: 

\begin{subequations}\label{eqn:F}
\begin{align}
\dot{\mathbf{F}} = 
\begin{pmatrix}
1.0\times 10^{-3} &  0&  0\\
0 & \ast & 0 \\
 0& 0 & \ast \\
\end{pmatrix}  \SI{}{\per\second},
\end{align}
\begin{align}
\mathbf{P} =
\begin{pmatrix}
\ast & \ast & \ast\\
\ast & 0 & \ast \\
 \ast & \ast & 0 \\
\end{pmatrix},
\end{align}

\end{subequations}

\noindent where $\ast$ symbol denotes unspecified components of the tensors. 



For the effective elastic modulus of Ren{\'e} 88DT, we ran simulations to approx.\ \SI{0.01}{\percent} total applied strain using purely elastic constitutive behavior of the material grid points. For the effective elastic modulus and yield strength of aluminum, we used elasto-plastic constitutive equations and MVEs were deformed to the total strain of \SI{0.04}{\percent}. For the plastic behavior, we adopted the power-law constitutive law with the following key model parameters: $\dot{\gamma}_0=$ \SI{0.001}{\second^{-1}}, $g_{0}=$ \SI{31}{\mega\pascal}, $g_{\infty}=$ \SI{63}{\mega\pascal}, and $h_0=$ \SI{75}{\mega\pascal} \cite{eisenlohr2013spectral}.

\subsection{Polycrystal graphs and graph neural networks} 
\label{sec:gnn} 

{\it{Polycrystal graphs}}. To allow GNNs learn microstructure--property relationships, we described 3D polycrystalline microstructures using undirected graphs. In these graphs, each node represents an individual grain, while the edges connect neighboring grains (with a shared boundary) taking periodicity in MVEs into account. Each grain-node in the graph was assigned a set of attributes, which included the grain volume (in voxels) and elements of either the quaternion vector describing the grain orientation (for SAGE-O models), or rotated tensor relevant to the effective property of interest (for AnisoGNN models). 
The AnsioGNN-C models for effective elastic modulus relied on the 21 elements of the full elastic stiffness tensor. The AnisoGNN-S models for modeling effective yield strength used (concatenated) 108 elements of the 12 Schmid tensors corresponding to 12 slip systems. Before being assigned to nodes as attributes, both the stiffness tensor and the Schmid tensors of each grain were rotated to the global frame using the grain orientations \cite{kalidindi1992crystallographic}. 

{\it{SAGE models}}. For SAGE-O models discussed above, we adopted the GNN architecture reported by Hestroffer et al.\ \cite{hestroffer2023graph}. The attributes (quaternions) of the nodes sequentially pass through a pre-processing fully-connected layer, two message-passing layers with the SAGE convolution \cite{hamiltonInductiveRepresentationLearning}, a global mean pooling layer, two post-processing fully-connected layers (all with ReLU activation except global mean pooling) followed by an output layer. All SAGE-O models were trained for 600 epochs. 

{\it{AnisoGNN models}}. For AnisoGNN models, we designed a simple GNN architecture with graph convolutional layer followed by global mean pooling and a post-processing fully-connected layer for output. The graph convolutional layer, based on a mean message passing scheme, first applies a linear transformation (with ReLU activation) to the node attributes, then aggregates attributes from neighboring nodes by averaging (mean aggregation). The fully connected layer of the convolutional layer linearly transforms the input from 22 (grain size and 21 stiffness tensor elements) to six features in the AnisoGNN-C models and from 109 (grain size and 108 elements of 12 Schmid tensors) to 19 in the AnisoGNN-S models. Global mean pooling consolidates transformed node attributes of each graph into a single feature vector. The network concludes with a layer that maps this pooled graph representation to the output. All AnisoGNN models were trained for 800 epochs. 



\bibliography{references}
\end{document}